\documentclass[twocolumn,showpacs,superscriptaddress,preprintnumbers,amsmath,amssymb]{revtex4}

\usepackage{graphicx}
\usepackage{dcolumn}
\usepackage{bm}

\makeatletter
\@ifundefined{textcolor}{}
{%
 \definecolor{BLACK}{gray}{0}
 \definecolor{WHITE}{gray}{1}
 \definecolor{RED}{rgb}{1,0,0}
 \definecolor{GREEN}{rgb}{0,1,0}
 \definecolor{BLUE}{rgb}{0,0,1}
 \definecolor{CYAN}{cmyk}{1,0,0,0}
 \definecolor{MAGENTA}{cmyk}{0,1,0,0}
 \definecolor{YELLOW}{cmyk}{0,0,1,0}
}

\usepackage{color}
\usepackage{dcolumn}
\usepackage{bm}

\makeatother

\begin{document}

\title{Exotic Topological States with Raman-Induced Spin-Orbit Coupling}

\author{Y. Deng}
\email{ygdeng2015@mail.tsinghua.edu.cn}
\affiliation{CAS Key Laboratory of Theoretical Physics, Institute of Theoretical Physics, Chinese Academy of Sciences, P.O. Box 2735, Beijing 100190, China}

\affiliation{State Key Laboratory of Low Dimensional Quantum Physics,Department of Physics, Tsinghua University, Beijing 100084, China}

\author{T. Shi}
\email{tshi@mpq.mpg.de}
\affiliation{Max-Planck-Institut f\"{u} Quantenoptik, Hans-Kopfermann-Strasse. 1, D-85748 Garching, Germany}

\author{H. Hu}
\affiliation{Centre for Quantum and Optical Science, Swinburne University of Technology, Melbourne 3122, Australia}

\author{L. You}
\affiliation{State Key Laboratory of Low Dimensional Quantum Physics,Department of Physics, Tsinghua University, Beijing 100084, China}
\affiliation{Collaborative Innovation Center of Quantum Matter, Beijing 100084, China}

\author{S. Yi}
\affiliation{CAS Key Laboratory of Theoretical Physics, Institute of Theoretical Physics, Chinese Academy of Sciences, P.O. Box 2735, Beijing 100190, China}
\affiliation{School of Physics, University of Chinese Academy of Sciences, P. O. Box 4588, Beijing 100049, China}

\date{\today}
\begin{abstract}
We propose a simple experimental scheme to realize simultaneously the one-dimensional spin-orbit coupling and the staggered spin flip in ultracold pseudospin-$1/2$ atomic Fermi gases trapped in optical lattices. In the absence of interspecies interactions, the system supports type-I and II Weyl semimetals in three-dimensional and gapped Chern insulators and gapless topological semimetal states in two-dimensional lattices. By turning on the $s$-wave interactions, a rich variety of gapped and gapless inhomogeneous topological superfluids can emerge. In particular, a gapped topological Fulde-Ferrell superfluid, which supports the chiral edge states at opposite boundaries with the same chirality, is predicted.
\end{abstract}


\maketitle
\section{Introduction}
Topological states that exhibit topologically protected excitations and gapless edge modes have attracted much attention in recent years~\cite{Xiao10,Hasan10,Qi11}. Spin-orbit (SO) coupling plays an essential role in such novel quantum states of solids, giving rise to the quantum spin-Hall effect~\cite{Kane05,Bernevig06,Hsieh08}, Majorana fermions~\cite{Mourik12,MTDeng,Das12}, and Weyl semimetals (WSMs)~\cite{Xusm15,Lusm15,Lvsm15,Soluyanovsm15}. Recently, the experimental realizations of the Raman-induced one-dimensional (1D) SO coupling in ultracold atomic gases \cite{Lin11,Wang12,Cheuk12} have offered a new paradigm for exploring a variety of topological states, including the gapped topological insulators or superfluids or gapless topological semimetals (tSMs) with unprecedented opportunities~\cite{zjy12,Dalibard11,Goldman2014,Zhai15,Sinha11,Deng12,Sato09,Jiang11,Gong11,Hu11,Zhai11,Xu11}. Ignoring collisional interactions, a rich variety of interesting topological phenomena has been experimentally addressed concerning single-particle physics of ultracold atoms in optical lattices, including the Su-Schrieffer-Heeger model~\cite{Atala13}, Hofstadter model~\cite{Aidelsburger13,Miyake13}, Haldane model~\cite{Jotzu14}, synthetic dimensions~\cite{Mancini15,Stuhl15}, and topological charge pumping~\cite{Nakajima16,Lohse16}, except for the gapless WSMs and tSMs which remain to be realized. When interactions are included, topological Fulde-Ferrell (FF) superfluids are theoretically predicted to appear in Rashba SO-coupled atomic Fermi gases in the presence of both in-plane and out-of-plane Zeeman fields~\cite{Zhang13,Qu13,Cao14,Xu14}. The inhomogeneous FF superfluid is characterized by Cooper pairs carrying a nonzero single-valued center-of-mass (c.m.) momentum. The atomic Rashba SO type remains a challenging task to synthesize, despite some recent theoretical proposals~\cite{Ruseckas05,Vaishnav08,Dalibard10,Campbell11,Xu11,Anderson12} and experimental advances in two-dimensional (2D) gases~\cite{Huang15,Wu15}. It is of great interest to answer if there exists a simple way to realize the exotic WSMs and tSMs for single particle spectra and observe inhomogeneous topological superfluids in atomic gases. An affirmative answer will significantly enhance our understanding of topological quantum matters and motivate the relevant studies in the condensed-matter community.

In this paper, we propose a readily implementable experimental scheme to realize simultaneously the 1D SO coupling and the staggered spin flip in pseudospin-$1/2$ atomic Fermi gases trapped in a three-dimensional (3D) cubic lattice. In the absence of the interatomic interactions, we show that the system supports WSMs. Particularly, for a reduced 2D lattice, our system realizes a nontrivial model Hamiltonian for a chiral $p+ip$ superconductor with an inversion asymmetric potential, which gives rise to the gapped Chern insulator (CI) and gapless tSMs. Furthermore, including effective attractive $s$-wave interactions in 2D lattices, a rich variety of gapped and gapless topological FF superfluid phases may appear. Of particular interest, we find a gapped topological FF superfluid, which hosts chiral edge states at opposite boundaries with the same chirality.

This paper is organized as follows. In Sec.~\ref{effham}, we introduce our model of Raman-induced spin-orbit coupling and derive the single-particle Hamiltonian for a pseudospin-$1/2$ atomic Fermi gases. In Sec.~\ref{weyl3D}, we study the phase transition between the type-I and -II Weyl semimetals in 3D lattices. In Sec.~\ref{topo2D}, we study the topological states and phase diagram in 2D lattices. In Sec.~\ref{topoFF}, we present the inhomogeneous topological superfluids in 2D lattices by utilizing the Green's-function method. In Sec.~\ref{expfea}, we discuss the experimental feasibility of our model. Finally, in Sec.~\ref{conclu}, we give a brief summary.

\section{Model and Hamiltonian}\label{effham}
\begin{figure}
\includegraphics[width=0.95\columnwidth]{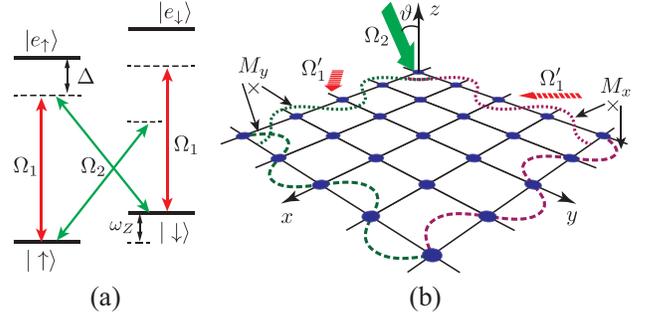} \protect\caption{(color online). (a) Level diagram of a four-level atomic system. (b) Proposed scheme for generating Raman-assisted hopping in the $xy$ lattice plane. The two lines with the same color linked to $M_{x,y}$ denote the process of staggered spin-flip Raman.}
\label{scheme}
\end{figure}

Our system consists of an ultracold gas of $N$ four-level fermionic atoms subjected to a bias magnetic field $\mathbf{\mathit{\mathbf{B}}}$ along the quantization $z$ axis. Figure~\ref{scheme}(a) displays the atomic level structure, and Fig.~\ref{scheme}(b) illustrates the laser configuration. Specifically, the atomic transition frequency from the electronic ground state $\left|\uparrow\right\rangle $ ($\left|\downarrow\right\rangle $) to the excited state $\left|e_{\uparrow}\right\rangle $ ($\left|e_{\downarrow}\right\rangle $) is $\omega_{a}$, and the magnetic quantum numbers of these electronic states satisfy $m_{\sigma}=m_{e_{\sigma}}$
($\sigma=\uparrow,\downarrow$) and $m_{\uparrow}=m_{\downarrow}+1$. The ground states $\left|\uparrow\right\rangle $ and $\left|\downarrow\right\rangle $ are split by the Zeeman shift $\hbar\omega_{Z}$ induced by the bias field. Moreover, we assume that atoms are deeply confined in a spin-independent red-detuned 3D cubic optical lattice ${\cal U}_{{\rm ol}}({\bf r})=-U_{{\rm ol}}\left[\cos^{2}(k_{L}x)+\cos^{2}(k_{L}y)+\gamma^2\cos^{2}(k_{L}z)\right]$ with the aspect ratio $\gamma$, the depth of the lattice $U_{{\rm ol}}$, and $k_{L}=\sqrt{2}\pi/\lambda$ with $\lambda$ being the wavelength of the Raman lasers. The lattice constant is $a=\pi/k_{L}$. To generate SO coupling, the transitions $|\sigma\rangle\leftrightarrow|e_{\sigma}\rangle$ are driven by a pair of $\pi$-polarized \emph{standing-wave} lasers with frequency $\omega_{L}$, which are detuned $\Delta=\omega_{a}-\omega_{L}$ from the atomic transitions. These two beams propagate along the directions ${\boldsymbol{e}}_{x}-{\boldsymbol{e}}_{y}$ and ${\boldsymbol{e}}_{x}+{\boldsymbol{e}}_{y}$ with Rabi frequencies $\Omega_{1}'\sin(k_{L}x-k_{L}y)$ and $i\Omega_{1}'\sin(k_{L}x+k_{L}y)$, respectively, where ${\boldsymbol{e}}_{\alpha}$ ($\alpha=x,y,z$) are the unit vectors along the $\alpha$ axis. Hence, the total Rabi frequency is $\Omega_{1}\left[\sin(k_{L}x)\cos(k_{L}y)+i\cos(k_{L}x)\sin(k_{L}y)\right]$ with $\Omega_{1}=\sqrt{2}e^{i\pi/4}\Omega_{1}'$. To complete the Raman process \cite{Wang12,Cheuk12}, the atoms are also illuminated by a $\sigma$-polarized (along the $y$ axis) plane-wave laser with frequency $\omega_{L}+\Delta\omega_{L}$ that propagates along the direction $\sin\vartheta{\boldsymbol{e}}_{x}+\cos\vartheta{\boldsymbol{e}}_{z}$  making an angle $\vartheta$ to the $z$ axis. The Rabi frequency of the plane-wave laser is $\Omega_{2}e^{i(\kappa_x x+\kappa_zz)}$, where $\Omega_{2}$ is real, $\kappa_x=\sqrt{2}k_{L}\sin\vartheta$, and $\kappa_z=\sqrt{2}k_{L}\cos\vartheta$. Compared with the earlier related Raman scheme for creating 1D SO coupling \cite{Wang12,Cheuk12}, the use of standing-wave $\pi$-polarized lasers leads to nontrivial staggered spin flip ${M}_{x}({\bf r})\hat{\sigma}_{x}\!+\!{M}_{y}({\bf r})\hat{\sigma}_{y}$ on the $xy$ plane, as we shall see below.

In the large-detuning limit, $|\Omega_{1,2}/\Delta|\ll1$, we adiabatically eliminate the excited states $|e_{\sigma}\rangle$, which leads to the Stark shifts $U_{1,2}=-\Omega_{1,2}^{2}/\Delta$ for both ground states $\left|\uparrow\right\rangle $ and $\left|\downarrow\right\rangle $ and Raman coupling $\Omega=-\Omega_{1}\Omega_{2}/\Delta$ between them.
Under the conditions $|\Omega/\Delta\omega_{L}|\ll1$, $|U_{1,2}/\Delta\omega_{L}|\ll1$, and $|\Delta\omega_{L}/\Delta|\ll1$, the off-resonant Raman process can be neglected due to the high-frequency prefactor $e^{\pm i2\Delta\omega_{L}t}$. After the gauge transformations $\left|\uparrow\right\rangle \!\rightarrow\! e^{-i(\kappa_x x+\kappa_z z)/2}\left|\uparrow\right\rangle $ and $\left|\downarrow\right\rangle \!\rightarrow \!e^{i(\kappa_x x+\kappa_z z)/2}\left|\downarrow\right\rangle$, the single-particle Hamiltonian reads~(see Appendix \ref{APPA})
\begin{eqnarray}
{\boldsymbol{h}}=\frac{({\mathbf{p}}-{\mathbf{A}})^{2}}{2M}\!+\!{M}_{x}({\bf r})\hat{\sigma}_{x}\!+\!{M}_{y}({\bf r})\hat{\sigma}_{y}\!-\!\frac{\delta}{2}\hat{\sigma}_{z}+{\cal U}({\bf r})I,\label{single-SO}
\end{eqnarray}
where $M$ is the atom mass; ${\mathbf{A}}=\hbar\kappa_x\hat{\sigma}_{z}\hat{\mathbf{x}}/2+\hbar\kappa_z\hat{\sigma}_{z}\hat{\mathbf{z}}/2$ is the vector potential, with $\kappa_x$ ($\kappa_z$) characterizing the strength of the 1D SO coupling along the $x$ ($z$) direction; $\hat{\sigma}_{x,y,z}$ are Pauli matrices; ${M}_{x}({\bf r})=\Omega\sin(k_{L}x)\cos(k_{L}y)$; ${M}_{y}({\bf r})=\Omega\cos(k_{L}x)\sin(k_{L}y)$; $\delta=\omega_{Z}+\Delta\omega_{L}$ is the two-photon detuning; ${\cal U}({\bf r})={\cal U}_{{\rm ol}}({\bf r})+{\cal U}_{1}({\bf r})$, with ${\cal U}_{1}({\bf r})=U_{1}[\sin^{2}(k_{L}x)\cos^{2}(k_{L}y)+\cos^{2}(k_{L}x)\sin^{2}(k_{L}y)]$, and $I$ is an identity matrix. Here we neglect a constant term, $-U_{2}+\delta/2$, in ${\cal U}({\bf r})$. We note that the Hamiltonian~(\ref{single-SO}) corresponds to an optical flux lattice with a nonzero net spatial magnetic flux and large synthetic magnetic field, as discussed in Appendix \ref{APPB}. A particular advantage of the present scheme is that it does not rely on the spin-dependent optical lattice for ultracold atomic gases~\cite{Cooper11,Deng14}.

For sufficiently strong lattice potential, the system enters the tight-binding regime such that the field operators of the atoms can be expanded in terms of the localized Wannier function $w_{\mathbf{j}}({\bf r})\equiv w({\bf r}-{\bf r}_{\mathbf{j}}$) of the lowest $s$ orbits, where ${\mathbf{j}}\equiv(m,n)$ is the
2D lattice-site index. We note that for $|U_{1}/U_{{\rm ol}}|\ll1$ and $|\Omega/U_{{\rm ol}}|\ll1$, $w({\bf r})$ is determined solely by the optical lattice potential ${\cal U}_{{\rm ol}}({\bf r})$. For convenience, we define three lattice unit vectors, ${\mathbf{1}}_{x}\!=\!(1,0,0)$, ${\mathbf{1}}_{y}\!=\!(0,1,0)$, and ${\mathbf{1}}_{z}\!=\!(0,0,1)$. Now, considering only the nearest-neighbor hoppings, the lattice Hamiltonian reads
\begin{align}
H_{0} &=\sum_{{\mathbf j},\zeta=\pm1}\left(t^{(R)}_{x}\, \zeta \hat{c}^\dag_{{\mathbf j},\uparrow}\hat{c}_{{\mathbf j}+\zeta {\mathbf 1}_{x},\downarrow}-it^{(R)}_y\,\zeta \hat{c}^\dag_{{\mathbf j},\uparrow}\hat{c}_{{\mathbf j}+\zeta{\mathbf 1}_{y},\downarrow}+{\rm H.c.}\right)\nonumber \\
& \quad-\sum_{{\mathbf j},\alpha=x,y}t_{\alpha}\left(\hat{c}^\dag_{{\mathbf j},\uparrow}{\mathcal R}_\alpha^{\uparrow\uparrow}\hat{c}_{{\mathbf j}+{\mathbf 1}_\alpha,\uparrow}-\hat{c}^\dag_{{\mathbf j},\downarrow}{\mathcal R}_\alpha^{\downarrow\downarrow}\hat{c}_{{\mathbf j}+{\mathbf 1}_\alpha,\downarrow}
+ {\rm H.c.}\right) \nonumber \\
& \quad-t_z\sum_{{\mathbf j}}\left(\hat{c}^\dag_{{\mathbf j},\uparrow}{\mathcal R}_z^{\uparrow\uparrow}\hat{c}_{{\mathbf j}+{\mathbf 1}_z,\uparrow}+\hat{c}^\dag_{{\mathbf j},\downarrow}{\mathcal R}_z^{\downarrow\downarrow}\hat{c}_{{\mathbf j}+{\mathbf 1}_z,\downarrow}
+ {\rm H.c.}\right) \nonumber \\
&\quad-\frac{\delta}{2} \sum_{{\mathbf j}}\left(\hat{n}_{{\mathbf j},\uparrow}-\hat{n}_{{\mathbf j},\downarrow}\right), \label{hamlat}
\end{align}%
where $\hat{c}_{{\mathbf{j}},\sigma}$ is the annihilation operator, $\hat{n}_{{\mathbf{j}},\sigma}\!=\!\hat{c}_{{\mathbf{j}},\sigma}^{\dag}\hat{c}_{{\mathbf{j}},\sigma}$, and $t_{\alpha}\!=\!\!-\!\int d{\bf r}\, w_{{\mathbf{i}}}^{*}({\bf r})\left[{\bf {p^{2}}}/(2m)\!+\!{\cal U}_{\text{ol}}({\bf r})\right]w_{{\mathbf{i}}+{\bf 1}_{\alpha}}({\bf r})$ is the spin-independent hopping matrix element, with $t_x=t_y\equiv t$. The matrix elements for Raman-assisted spin-flip hopping $t^{(R)}_{x,y}\!=\!\Omega\int d{\bf r}w_{{\mathbf{i}}}^{*}({\bf r})|M_{x,y}({\bf r})|w_{{\mathbf{i}}+{\bf 1}_{x,y}}({\bf r})$
are the same along the $x$ and $y$ directions, i.e., $t^{(R)}_{x}=t^{(R)}_{y}\equiv t_{0}$. Note that $M_x$ ($M_y$) does not contribute to spin-flip hopping along the $y$ ($x$) axis. The plane-wave laser propagating in the $xz$ plane introduces a Peierls substitution ${\mathcal{R}}_{\alpha}=\exp[-(i/\hbar)\int_{{\bf r}_{\mathbf{j}}+a{\boldsymbol{e}}_{\alpha}}^{{\bf r}_{\mathbf{j}}}{\bf A}\cdot d{\bf l}]$ along the ${\boldsymbol{e}}_{\alpha}$ direction with matrix elements denoted as ${\cal R}_{\alpha}^{\sigma\sigma'}$. Explicitly, we have ${\mathcal{R}}_{x,z}=\exp(i\phi_{x,z}\hat{\sigma}_{z})$ and ${\mathcal{R}}_{y}={I}$, where the Peierls phase $\phi_{x,z}=\kappa_{x,z} a/2$ is controllable through the angle $\vartheta$. To obtain Eq.~(\ref{hamlat}), a gauge transformation $\hat{c}_{{\mathbf{j}},\downarrow}^{\dag}\rightarrow(-1)^{n+m+1}\hat{c}_{{\mathbf{j}},\downarrow}^{\dag}$~\cite{Liu14}
is applied to eliminate the staggered factor in the spin-flip hopping.

In momentum space, the Hamiltonian (\ref{hamlat}) reduces to
\begin{eqnarray}
H_{0}({\mathbf{k}})=\sum_{{\mathbf{k}},\sigma\sigma'}\hat{c}_{{\mathbf{k}}\sigma}^{\dag}\Big[\epsilon({\mathbf{k}})I+\sum_{\alpha=x,y,z}d_{\alpha}({\mathbf{k}})\hat{\sigma}_{\alpha}\Big]_{\sigma\sigma'}\hat{c}_{{\mathbf{k}}\sigma'},\label{k-single}
\end{eqnarray}
where ${\mathbf{k}}=(k_{x},k_{y},k_{z})$ is in the first Brillouin zone (FBZ), $d_{x}({\mathbf{k}})=2t_{0}\sin(k_{y}a)$, $d_{y}({\mathbf{k}})=-2t_{0}\sin(k_{x}a)$, $d_{z}({\mathbf{k}})=-\delta/2-2t[\cos\phi_x\cos(k_{x}a)+\cos(k_{y}a)] +2t_z\sin\phi_z \sin(k_{z}a)$, and $\epsilon({\mathbf{k}})=2t\sin\phi_x\sin(k_{x}a)-2t_z\cos\phi_z\cos(k_{z}a)$. Here the spin-flip terms $d_{x,y}({\mathbf{k}})\hat{\sigma}_{x,y}$ preserve the time-reversal (TR) symmetry. The gauge potential induced dispersion $\epsilon({\mathbf{k}})$ breaks the TR and inversion ($\mathcal{P}$) symmetries simultaneously, which, as will be shown, is essential to type-II WSM~\cite{Soluyanovsm15} in 3D lattices and topological tSM phases and FF superfluids in 2D lattices. Finally, we remark that all control parameters, $\vartheta$, $\delta/t$, $t_{z}/t$, and $t_{0}/t$, are independently tunable.

\begin{figure}
\includegraphics[width=0.85\columnwidth]{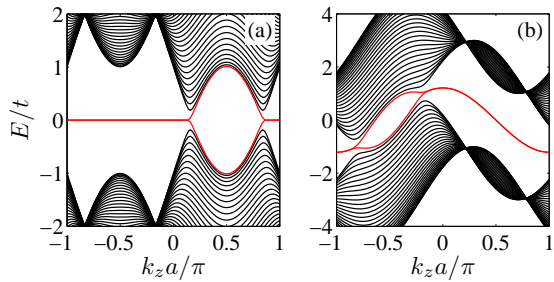}
\protect\caption{(color online). Energy spectra of (a) type-I WSM with $\vartheta=\pi/4$ and (b) type-II WSM with  $\vartheta=0$ for $k_xa=0$, $t_0=t_z=t$, and $\delta/t=-2$. The surface states are denoted by the red solid lines.} \label{spc-weyl}
\end{figure}

\section{Weyl semimetals in 3D lattices}\label{weyl3D}

Diagonalizing Hamiltonian~(\ref{k-single}), we obtain two energy bands: $E_{\pm}({\bf k})=\epsilon({\mathbf{k}})\pm|{\mathbf{d}}({\mathbf{k}})|$. It can be shown that, by tuning control parameters, we can find at least a pair of gapless points defined by the solutions of $|{\mathbf{d}}({\mathbf{k}})|=0$ (see Appendix \ref{APPA}). Without loss of generality, we consider the pair located at ${\bf K}_{+}=(0,0,k_w)$ and ${\bf K}_{-}=(0,0,\pi/a-k_w)$, with $k_wa=\sin^{-1}[(\delta/4 + t\cos\phi_x+t)/(t_z\sin\phi_z)]$. Expanding Eq.~(\ref{k-single}) in the vicinity of ${\mathbf K}_{\pm}$, the effective Hamiltonian, after dropping a constant, takes the form
\begin{eqnarray}
H_{\pm}(\tilde{\mathbf k})=v_0(\tilde{\mathbf k}){I} + v_x \tilde k_x\hat{\sigma}_y+v_y\tilde k_y \hat{\sigma}_x \pm v_z\tilde k_z \hat{\sigma}_z, \label{Weyl-ham}
\end{eqnarray}
where $\tilde{\mathbf k}$ is the wave vector with respect to ${\mathbf K}_{\pm}$, $v_x=-2t_0a $, $v_y=2t_0a$, $v_z=2ta\sin\phi_z\cos(k_wa)$, and $v_0({\bf k})=v_{0}^{(z)} \tilde k_z+v_{0}^{(x)}\tilde k_x$ with $v_{0}^{(z)}=2t_za\cos\phi_z\sin(k_wa)$ and $v_{0}^{(x)}=2ta\sin\phi_x$. The linear energy dispersion for the momenta along all directions clearly proves that ${\mathbf K}_{\pm}$ are Weyl points. The topology of the Weyl points is determined by the first Chern number, $C = (2\pi)^{-1} \oint d {\bf S}\cdot {\Omega_-}({\bf k})$, an surface integral of the Berry curvature ${\Omega_-}({\bf k})= {{\bf d}}/(2{d^3})$~\cite{Xiao10} over the surface enclosing the Weyl point. We find that the Weyl points ${\mathbf K}_{\pm}$ indeed have opposite chiralities as their corresponding Chern numbers are $C_{\pm}=\pm{\rm{sgn}}(v_xv_yv_z)$, which is in contrast to the Weyl points proposed in Ref.~\cite{Xusm16} that possess  topological charge.

Interestingly, the $v_{0}(\tilde{\mathbf k})I$ term in Eq.~(\ref{Weyl-ham}) which tilts the spectrum allows us to further classify the Weyl point based on the classification of the Fermi surface~\cite{Soluyanovsm15}. In fact, it can be readily shown that, when $\left|v_{0}^{(x)}/v_x\right|<1$ and $\left|v_{0}^{(z)}/v_z\right|<1$, the Weyl point (WP) has a pointlike Fermi surface and is classified as type I (standard). Otherwise, it is a type-II Weyl point, for which the Fermi surface has both electron and hole pockets due to the highly tilted spectrum. The difference between these two types of WPs is also revealed by the corresponding surface states. In Fig.~\ref{spc-weyl}, we plot the typical band spectra calculated with the open boundary condition along the $y$ axis for type-I and -II WPs. Both types of Weyl semimetals support surface states which connect two WPs. However, unlike the type-I Weyl point, the $z$ components of the velocities $\partial E/\partial{k_z}$ possess the same sign for the type-II Weyl point. Finally, we point out that, in our model, the transition between type-I and -II Weyl points (Lifshitz transition) can be easily induced by tuning $\vartheta$, the tilt angle of the plane-wave laser.

\section{Topological states in 2D lattices}\label{topo2D}
\begin{figure}
\includegraphics[width=0.99\columnwidth]{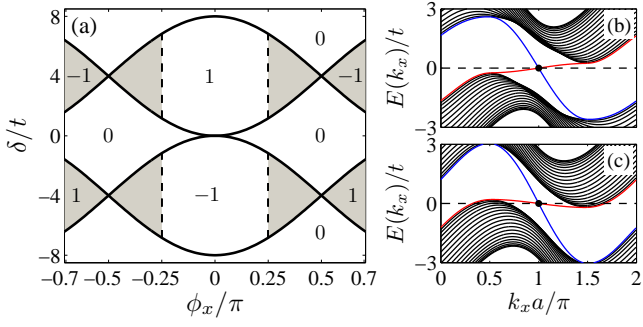} \protect\caption{(color online). (a) Phases of the Hamiltonian~(\ref{k-single}) on the $\phi_x\delta$ plane. Equations (\ref{cricond}) are plotted as solid lines and the numbers represent the Chern numbers of the different phases. Dashed lines denote the boundaries between CI and tSM phases for $t_{0}/t=0.71$ with shaded regions marking the tSM phase. (b) and (c) show the band spectra for $\phi_x/\pi=0.2$ (CI phase) and $0.3$ (tSM phase), respectively, for the parameters $t_{0}/t=0.71$ and $\delta/t=4$.}
\label{sphase}
\end{figure}

Our model also hosts a rich variety of 2D topological states. To see this, we assume that the lattice potential long the $z$ direction is so strong that the hopping along the $z$ direction is prohibited ($t_{z}=0$), which reduces our system to 2D. In the reduced 2D Hamiltonian (see Appendix \ref{APPA}), $\epsilon({\mathbf k})$ becomes an odd function of $k_{x}$ such that it plays a role similar to that of the layer-asymmetric stain in bilayer graphene~\cite{Crosse14}, which, as will be shown below, is essential to tSMs. To search for the topological states, we again consider the gapless points, defined by the vanishing direct bulk gap $E_{g}^{(d)}=2\min[|{\mathbf{d}}({\mathbf{k}})|]$. It can easily be shown that the condition leads to four curves,
\begin{eqnarray}
\frac{\delta}{t}=\pm4\left(\cos\phi_x\pm1\right),\label{cricond}
\end{eqnarray}
which, as plotted in Fig.~\ref{sphase}(a), divide the $\phi_x\delta$ parameter plane into three regions associated with different Chern numbers. Therefore, these curves indeed define the critical point for the topological phase transition. However, it should be noted that even in a topologically nontrivial region, the system is not necessarily a Chern insulator due to the inversion asymmetry of $\epsilon({\mathbf{k}})$. For a complete characterization of a state, we also need to consider the indirect bulk gap $E_{g}^{(i)}=\min[E_{+}({\boldsymbol{k}})]-\max[E_{-}({\boldsymbol{k}})]=2\min[E_{+}({\mathbf{k}})]$. In the topologically nontrivial regions, the system belongs to the tSM phase if the indirect gap is closed; otherwise, it is a CI. Analytically, we find that the condition for the tSM phase is $|\sin\phi_x|>t_{0}/t$, which is equivalent to $|v_{0}^{(x)}/v_x|>1$. Hence the tSM phase vanishes when $t_{0}\geq t$. Although the critical point corresponding to the topological phase transitions {[}Eq.~(\ref{cricond}){]} is independent of $t_{0}$, the phase boundary between CI and tSM can be tuned by varying $t_{0}$.

To gain more insight into the topological states, we consider the edge modes of the system by imposing a hard-wall confinement along the $y$ direction. Figures~\ref{sphase}(b) and \ref{sphase}(c) show the typical energy spectra $E(k_{x})$ in the CI and tSM phases, respectively, and confirms that both topological phases support edge modes. Unlike the counter-propagating edge modes in the CI phases, the inversion asymmetry of $\epsilon(\mathbf{k})$ in the tSM phase gives rise to the same chirality for both edge modes. The velocities, $\partial E(k_{x})/\partial k_{x}$, of the edge states at different boundaries have the same sign. Remarkably, the chirality of the edge modes is also tunable by negating the Peierls phase $\phi_x$, which changes $\epsilon({\mathbf{k}})$ to $-\epsilon({\mathbf{k}})$. We remark that the position of the TR-invariant point depends on only the sign of $\delta$, i.e., $k_{x}a=0$ ($\pi$) for $\delta<0$ ($>0$).

\begin{figure}
\includegraphics[width=0.99\columnwidth]{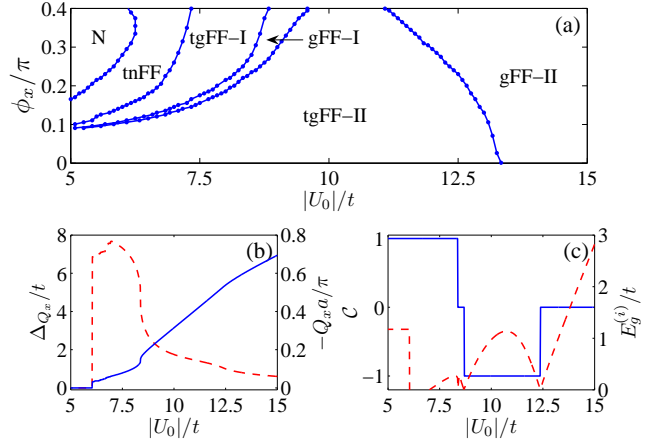} \protect\caption{(color online). (a) Phase diagram in the $U_{0}\phi_x$ parameter plane. (b) $|U_{0}|$ dependence of $\Delta_{Q_{x}}$ (solid line) and $Q_{x}$ (dashed line). (c) $|U_{0}|$ dependence of ${\cal C}$ (solid line) and $E_{g}^{(i)}$ (dashed line).}
\label{mphase}
\end{figure}

\section{Topological Fulde-Ferrell superfluids}\label{topoFF}
We now turn to the zero-temperature superfluids in 2D lattices including an attractive $s$-wave interaction $H_{\text{int}}=(U_{0}/\mathcal{S})\sum_{{\bf kk'q}}\hat{c}_{{\bf q/2+k},\uparrow}^{\dag}\hat{c}_{{\bf q/2-k},\downarrow}^{\dag}\hat{c}_{{\bf q/2-k'},\downarrow}\hat{c}_{{\bf q/2+k'},\uparrow}$, which
is invariant under spin rotation, where $\mathcal{S}$ is the number of lattice sites and $U_{0}$ ($<0$) is the interaction strength. We note that although the attractive $s$-wave interaction may lead to the standard Bardeen-Cooper-Schrieffer (BCS) superfluid, the ${\cal P}$ symmetry breaking $\epsilon({\mathbf{k}})$ often induces the asymmetric energy bands that favor a nonzero pairing momentum. In the presence of the FF superfluid, the order parameter $\Delta_{{\bf Q}}\equiv(U_{0}/\mathcal{S})\sum_{{\bf k}}\langle\hat{c}_{{\bf Q/2-k},\downarrow}\hat{c}_{{\bf Q/2+k},\uparrow}\rangle$
possesses a nonzero c.m. momentum ${\bf Q}$. In the Nambu representation, the mean-field Hamiltonian is
\begin{align}
H=\frac{1}{2}\sum_{{\bf k}}\hat{\Psi}_{{\bf k,Q}}^{\dag}{\mathcal{H}}_{\text{BdG}}\hat{\Psi}_{{\bf k,Q}}-\frac{\mathcal{S}}{U_{0}}|\Delta_{{\bf Q}}|^{2}+\sum_{{\bf k}}\xi_{{\bf k}},
\end{align}
where $\hat{\Psi}_{{\bf {k,Q}}}=(\hat{c}_{{\bf {Q/2+k},\uparrow}},\hat{c}_{{\bf {Q/2+k},\downarrow}},\hat{c}_{{\bf {Q/2-k},\downarrow}}^{\dagger},-\hat{c}_{{\bf {Q/2-k},\uparrow}}^{\dagger})^{T}$ is the Nambu operator, and $\xi_{{\bf k}}=\epsilon({\mathbf{k}})-\mu$, with $\mu$ being the chemical potential. By diagonalizing the Bogoliubov-de
Gennes (BdG) Hamiltonian, ${\mathcal{H}}_{\text{BdG}}({\bf k}){\psi}_{\eta,{\bf k}}^{\nu}=E_{\eta,{\bf k}}^{\nu}{\psi}_{\eta,{\bf k}}^{\nu}$, we obtain the eigenenergies $E_{\eta,{\bf k}}^{\nu}$ and the wave functions ${\psi}_{\eta,{\bf k}}^{\nu}$ of the Bogoliubov quasiparticles, where $\nu=\pm$ represents the particle ($+$) and hole ($-$) bands, and $\eta=1$ and $2$ denote, respectively, the upper and lower helicity branches~\cite{Cao14}. Here, because ${\mathcal{H}}_{\text{BdG}}({\bf k})$ preserves the inherent particle-hole (PH) symmetry but breaks the TR symmetry, it belongs to symmetry class $D$ according to the classification of Altland and Zirnbauer~\cite{AZ,Ludwig08}. In addition, the fact that ${\rm Tr}[\mathcal{H}_{{\rm BdG}}({\mathbf{k}})]=4\epsilon({\mathbf{k}})\cos(Q_{x}a/2)$ indicates that ${\mathbf{Q}}$ has to be along the $x$ direction, i.e., ${\bf Q}=Q_{x}{\mathbf{e}}_{x}$, if a FF superfluid exists. This result is also confirmed by our numerical simulations. Particularly, we find that a standard BCS superfluid ($Q_{x}=0$) emerges when $\phi_x=0$. The topological properties of the system are characterized by the Chern number computed using the hole branch, i.e., ${\cal {C}}=(1/2\pi)\sum_{\eta=1,2}\int_{{\bf {k}\in{\text{FBZ}}}}d{\bf k}{\Omega}_{\eta}^{-}({\bf k})$~\cite{TKNN},
where ${\Omega}_{\eta}^{\nu}({\bf k})=\nabla_{{\bf k}}\times{\cal A}_{\eta}^{\nu}({\bf k})$ is the Berry curvature with ${\cal A}_{\eta}^{\nu}({\bf k})=i\hbar\langle{\psi}_{\eta,{\bf k}}^{\nu}|\nabla_{{\bf k}}|{\psi}_{\eta,{\bf k}}^{\nu}\rangle$.
Analogously, the indirect bulk gap between the particle and hole branches, $E_{g}^{(i)}\!=\!2\min(E_{2,{\bf k}}^{-})$, is needed for a complete characterization of a state.

For the numerical method, instead of minimizing the free energy using numerical differentiation~\cite{Zhang13,Qu13,Cao14,Xu14}, we perform the derivatives analytically via the Green's-function method, which generates $\Delta_{Q_{x}}$, $\mu$, and ${Q_{x}}$ with high precision in Appendix \ref{APPC}. Furthermore, in all calculations, we take, without loss of generality, $t_{0}=t$, $\delta=4t$, and $n\equiv N/\mathcal{S}=0.8$. Figure~\ref{mphase}(a) summarizes the quantum phases in the $U_{0}\phi_x$ parameter plane. Here, a superfluid phase with a nonzero pairing momentum is denoted by ``FF". A state with $\Delta_{Q_{x}}/t<1.0\times10^{-2}$ is considered as a normal state and hence labeled by ``N". A topologically nontrivial state ($|{\cal C}|=1$ in this work) is denoted by ``t". Finally, ``g" and ``n" denote the gapped and gapless states, respectively.

To understand these phases, fixing $\phi_x/\pi=0.25$, we plot the FF momentum $Q_x$ and order parameter $\Delta_{Q_x}$ as functions of attractive interaction $|U_0|$ in Fig.~\ref{mphase}(b). Consequently, we find that $\Delta_{Q_{x}}=0$ at the small $|U_{0}|$ limit and it increases as $|U_{0}|$ increases. On the other hand, $|Q_{x}|$ decreases with increasing $|U_{0}|$. Moreover, we find that the chemical potential $\mu$ is roughly unchanged with fixing the particle density number $n$. Figure~\ref{mphase}(c) displays the Chern number and indirect bulk gap as a function of $|U_{0}|$. As we can see, the system also experiences various topological phase transitions as one increases $|U_{0}|$. Associated with the sudden changes of the Chern number, the indirect bulk gap is closed and reopened at the phase boundaries.

\begin{figure}
\includegraphics[width=0.99\columnwidth]{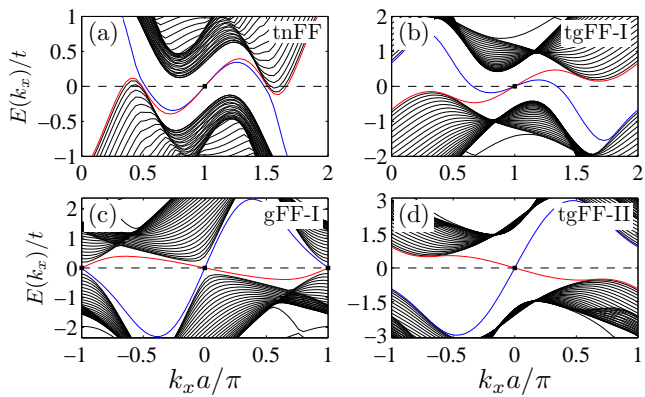} \protect\caption{(color online). (a)-(d) Quasiparticle spectra for $|U_{0}|/t=6.67$, $8.17$, $8.5$, and $10$ with $\phi_x/\pi=0.25$, respectively. The corresponding phase is denoted. The black dots mark the TR-invariant points.}
\label{sfphase}
\end{figure}

To further characterize these phases, we plot the quasiparticle spectrum $E(k_{x})$ of each phase in Figs.~\ref{sfphase}(a)-\ref{sfphase}(d). For small $|U_{0}|$, the system falls into the gapless tnFF phase where two edge states possess the same chirality, as shown in Fig.~\ref{sfphase}(a), in analogy to that of the tSM phase [see Fig. \ref{sphase}(c)]. With increasing $|U_{0}|$, the increased $\Delta_{Q_{x}}$ opens up the indirect bulk gap, signaling that the system is in the tgFF phase. Because its quasiparticle spectrum Fig.~\ref{sfphase}(b) shows that the left and right edge modes have the same chirality around the TR-invariant point, we denote this phase as tgFF-I to distinguish it from the tgFF-II state whose edge modes have opposite chirality {[}Fig.~\ref{sfphase}(d){]}. As $|U_{0}|$ is increased further, the system enters the gFF-I phase and the Chern number jumps to zero. Although it is a topologically trivial state, it has two TR-invariant points and edge modes, as identified in Fig.~\ref{sfphase}(c) \cite{gFF-I}. For large $U_0$, it naturally expects a trivial topological gFF-II phase without any edge modes in contrast to the gFF-I phase. We also numerically verified that each zero-energy mode at the TR-invariant points in Fig.~\ref{sfphase}(a)-\ref{sfphase}(d) corresponds to a Majorana fermion.

\section{Experimental feasibility}\label{expfea}
In principle, the proposed scheme should be applicable to most alkali-metal atoms~\cite{Wang12,Cheuk12}. Here, as an example, we discuss in detail how to implement our scheme in ${}^{40}$K atoms based on the parameters used in Ref.~\cite{Wang12}. First, two ground states can be chosen as $|\uparrow\rangle=|F=9/2,m_F=9/2\rangle$ and $|\downarrow\rangle=|F=9/2,m_F=7/2\rangle$. The wave length of Raman lasers may be taken as $\lambda=773\,{\rm nm}$ which is longer than the $D_{1}$ line. Consequently, the Raman lasers are red detuned with detuning $\Delta/(2\pi)=1.64\,{\rm THz}$. The recoil energy is then $E_L=\hbar^2k_L^2/2m=h\times 4.15\,{\rm kHz}$, and Raman coupling strength is $\Omega=0.5E_L$. Next, we choose a bias magnetic field of $B=31\,{\rm G}$, which leads to a Zeeman shift of $\omega_Z/(2\pi)=10.28\,{\rm MHz}$. In addition, under this magnetic field, the quadratic Zeeman shift can be as large as $41 E_L$, which is much larger than the Raman coupling strength such that the selected states are well separated from other hyperfine states in the $F=9/2$ manifold. Finally, the frequency difference between two Raman beams $\Delta\omega_{L}\approx\omega_{Z}$ is introduced as a tunable parameter.

As to the experimental detection, the predicted topological states can be detected using demonstrated capabilities by measuring the closing and opening of the bulk gap via the Landau-Zener transition~\cite{Jotzu14}, the Chern number of bands~\cite{Aidelsburger15}, and the Bloch band topology~\cite{Duca15,Li16,Flaschner16}.

\section{Conclusions}\label{conclu}
We have proposed an experimental scheme to generate 1D SO coupling in pseudospin-$1/2$ Fermi gases trapped in optical lattices. It has been shown that our system supports the exotic type-II WSMs and tSMs in single-particle spectra of 3D and 2D lattices, respectively. In the presence of attractive interaction, it also hosts the gapped and gapless topological superfluids with FF pairings in 2D lattices. The predicted exotic phases can exist over a wide range of control parameters in experimental systems currently available.

\section*{ACKNOWLEDGMENTS}
The authors thank X. Wan and K. Yang for valuable discussions. L.Y. is supported by the MOST (Grant No. 2013CB922004) of the National Key Basic Research Program of China and by NSFC (Grants No. 91121005 and No. 91421305). S.Y. is supported by the NSFC (Grants No. 11434011, No. 11421063, and No. 11674334). H.H. is supported by the ARC DP projects (DP140103231). T.S. acknowledges support from the European Union Integrated project Simulators and Interfaces with Quantum Systems (SIQS). Y.D. acknowledges support from the project funded by the China Postdoctoral Science Foundation and NSFC (Grant No. 11604178).

\appendix
\begin{widetext}
%

\section{Single-particle Hamiltonian}
\label{APPA}

Here let us first derive the single-atom Hamiltonian for the given level diagram and laser configuration in Fig. 1 of the main text. To this end, we first note that the total Rabi frequency corresponding to the transitions $|\sigma\rangle\leftrightarrow|e_{\sigma}\rangle$ that are driven by a pair of $\pi$-polarized standing-wave lasers is
\begin{align}
\Omega_1({\bf r})&=\Omega_1'\left[\sin(k_Lx-k_Ly) +i\sin(k_Lx+k_Ly)\right]\nonumber \\
&=\Omega_1'\left[\sin(k_Lx)\cos(k_Ly) - \cos(k_Lx)\sin(k_Ly) + i\sin(k_Lx)\cos(k_Ly) + i\cos(k_Lx)\sin(k_Ly)\right]\nonumber \\
&=\Omega_1'(1+i)\left[\sin(k_Lx)\cos(k_Ly) +i\cos(k_Lx)\sin(k_Ly)\right]
\nonumber \\
&=\Omega_1\left[\sin(k_Lx)\cos(k_Ly) +i\cos(k_Lx)\sin(k_Ly)\right].
\end{align}
As a remainder, the effective Rabi frequency for the other laser light is $\Omega_{2}e^{i(\kappa_x x+\kappa_zz)}$. Now, from the level diagram, it can be read out that, under the rotating-wave approximation, the Hamiltonian for the internal states of an atom is ($\hbar=1$)
\begin{align}
{\boldsymbol h}_{\rm in} &= \left[\Omega_1^*({\bf r})e^{i\omega_L
t}\left(\hat{b}_{\uparrow}^{\dag}\hat{e}_{\uparrow}+\hat{b}_{\downarrow}^{\dag}\hat{e}_{\downarrow}\right)+
\Omega^*_2({\bf r})e^{i(\omega_L+\Delta
\omega_L)t}\hat{b}_{\uparrow}^{\dag}\hat{e}_{\downarrow}+\Omega^*_2({\bf r})e^{i(\omega_L+\Delta
\omega_L)t}\hat{b}_{\downarrow}^{\dag}\hat{e}_{\uparrow} + {\rm H.c.}\right]\nonumber\\
&\quad +  \omega_Z\hat{b}^\dag_{\downarrow} \hat{b}_{\downarrow}+ \omega_a \hat{e}^\dag_{\uparrow} \hat{e}_{\uparrow} + (\omega_a+\omega_Z') \hat{e}^\dag_{\downarrow} \hat{e}_{\downarrow}, \label{SM-single}
\end{align}
where $\hat{b}_{\sigma=\uparrow,\downarrow}$ and $\hat{e}_{\sigma=\uparrow,\downarrow}$ are, respectively, the annihilation operators for ground and excited states and $\omega_{Z}'$ is the Zeeman shift of the excited states. By introducing the rotating frame that is defined by the unitary transformation
\begin{align}
\widetilde {\mathcal U}=\exp\left\{-i\left[-\Delta\omega_L \hat{b}^\dag_{\downarrow} \hat{b}_{\downarrow}  + \omega_L \hat{e}^\dag_{\uparrow} \hat{e}_{\uparrow}+ (\omega_L-\Delta\omega_L) \hat{e}^\dag_{\downarrow} \hat{e}_{\downarrow}\right] t\right\},
\end{align}
the internal-state Hamiltonian Eq.~(\ref{SM-single}) reduces to
\begin{align}
{\boldsymbol h}_{\rm in} &\rightarrow\widetilde {\mathcal U}^\dag {\boldsymbol h}_{\rm in}\widetilde {\mathcal U}- i \widetilde {\mathcal U}^\dag\frac{\partial}{\partial t}\widetilde {\mathcal U} \nonumber \\
&= \left[\Omega_1^*({\bf r})\left(\hat{b}_{\uparrow}^{\dag}\hat{e}_{\uparrow}+\hat{b}_{\downarrow}^{\dag}\hat{e}_{\downarrow}\right)+\Omega^*_2({\bf r})\hat{b}_{\downarrow}^{\dag}\hat{e}_{\uparrow}+\Omega^*_2({\bf r})e^{i2\Delta
\omega_Lt}\hat{b}_{\uparrow}^{\dag}\hat{e}_{\downarrow}+{\rm H.c.}\right]+  \delta\hat{b}^\dag_{\downarrow} \hat{b}_{\downarrow} + \Delta \hat{e}^\dag_{\uparrow} \hat{e}_{\uparrow} + (\Delta+\Delta\omega_L+\omega_Z') \hat{e}^\dag_{\downarrow} \hat{e}_{\downarrow}.
\end{align}

To eliminate the excited states, we first write down the Heisenberg equation of motion for the atomic operators
\begin{subequations}
\begin{align}
i\frac{\partial{\hat b}_{\uparrow}}{\partial t} &= \Omega_1^*({\bf r})\hat{e}_{\uparrow} + \Omega_2^*({\bf r})e^{i2\Delta
\omega_Lt}\hat{e}_{\downarrow},\\
i\frac{\partial{\hat b}_{\downarrow}}{\partial t} &= \delta\hat{b}_{\downarrow}+ \Omega_1^*({\bf r})\hat{e}_{\downarrow} + \Omega_2^*({\bf r})\hat{e}_{\uparrow},\\
i\frac{\partial{\hat e}_{\uparrow}}{\partial t} &= (\Delta- i\frac{\gamma}{2})\hat{e}_{\uparrow}+ \Omega_1({\bf r})\hat{b}_{\uparrow} + \Omega_2({\bf r})\hat{b}_{\downarrow},\\
i\frac{\partial{\hat e}_{\downarrow}}{\partial t} &= (\Delta+\Delta\omega_L+\omega_Z'- i\frac{\gamma}{2})\hat{e}_{\downarrow}+ \Omega_1({\bf r})\hat{b}_{\downarrow} + \Omega_2({\bf r})e^{-i2\Delta
\omega_Lt}\hat{b}_{\uparrow}, \label{EOM}
\end{align}
\end{subequations}
where we have formally included the spontaneous emission rate $\gamma$ for excited states. Now, in the large detuning limit, $|\Omega_{1,2}/\Delta|\ll1$, $|\delta/\Delta|\ll1$, and$|\gamma/\Delta|\ll1$, the excited states can be adiabatically eliminated by setting $i\dot{\hat{e}}_{\uparrow,\downarrow}=0$, which yields
\begin{subequations}
\begin{align}
{\hat e}_{\uparrow} &\approx -\frac{\Omega_1({\bf r})\hat{b}_{\uparrow} + \Omega_2({\bf r})\hat{b}_{\downarrow}}{\Delta}, \\
{\hat e}_{\downarrow} &\approx -\frac{\Omega_1({\bf r})\hat{b}_{\downarrow} + \Omega_2({\bf r})e^{-i2\Delta
\omega_Lt}\hat{b}_{\uparrow}}{\Delta}. \label{EOM1}
\end{align}%
\end{subequations}
Inserting these expression for ${\hat{e}}_{\uparrow,\downarrow}$  into the dynamical equations of ${\hat{b}}_{\uparrow,\downarrow}$, we find
\begin{subequations}
\begin{align}
i\frac{\partial{\hat b}_{\uparrow}}{\partial t} &=
-\frac{1}{\Delta}\left(\left[|\Omega_1({\bf r})|^2 +|\Omega_2({\bf r})|^2 \right]\hat{b}_{\uparrow}+
\left[\Omega^*_1({\bf r})\Omega_2({\bf r}) +\Omega^*_2({\bf r})\Omega_1({\bf r}) e^{i2\Delta\omega_Lt}\right] \hat{b}_{\downarrow}\right), \\
i\frac{\partial{\hat b}_{\downarrow}}{\partial t} &= \delta{\hat b}_{\downarrow}
-\frac{1}{\Delta}\left(\left[|\Omega_1({\bf r})|^2 +|\Omega_2({\bf r})|^2\right]\hat{b}_{\downarrow} +\left[\Omega^*_2({\bf r})\Omega_1({\bf r}) +\Omega^*_1({\bf r})\Omega_2({\bf r}) e^{-i2\Delta\omega_Lt}\right]\hat{b}_{\uparrow}\right).
\end{align}
\end{subequations}
The effective Hamiltonian for the ground states can then be easily read out as
\begin{eqnarray}
{\boldsymbol h}_{\rm in} = \begin{pmatrix}
-{\delta}/{2} & [{M}_{x}({\bf r}) -i {M}_{y}({\bf r})]e^{i(\kappa_x x+\kappa_z z)} \\
[{M}_{x}({\bf r}) +i {M}_{y}({\bf r})]e^{-i(\kappa_x x+\kappa_z z)}  & {\delta}/{2}\end{pmatrix}+{\cal U}_{1}({\boldsymbol\rho}) I,
\label{smsin1}
\end{eqnarray}
where the off-resonant Raman terms with the high-frequency prefactor $e^{\pm i2\Delta\omega_{L}t}$ have been neglected under the condition $|\Omega/\Delta\omega_{L}|\ll1$. Finally, incorporating the center-of-mass motion, the single-atom Hamiltonian becomes
\begin{align}
{\boldsymbol h}=\left[\frac{{\mathbf p}^{2}}{2M}+{\cal U}_{\rm ol}({\bf r})\right]\hat I+{\boldsymbol h}_{\rm in},
\end{align}
which, after the gauge transformations
$\left|\uparrow\right\rangle \!\rightarrow\! e^{-i(\kappa_x x+\kappa_z z)/2}\left|\uparrow\right\rangle $ and $\left|\downarrow\right\rangle \!\rightarrow \!e^{i(\kappa_x x+\kappa_z z)/2}\left|\downarrow\right\rangle$, gives rise to the Hamiltonian Eq. (1) in the main text.

\subsection*{Gapless points of the energy band in the 3D lattice}
From Hamiltonian (\ref{k-single}) in the main text, the lower and upper bands touch when $|{\mathbf d}({\bf k})|=0$. It can be easily derived that the locations of the gapless points in the momentum space are
\begin{enumerate}
\item $\displaystyle{(0,0,k_w)}$ and $\displaystyle{\left(0,0,\frac{\pi}{a}-k_w\right)}$, where $k_wa=\sin^{-1}c_{1}$ if $\displaystyle{|c_{1}|\equiv\left|\frac{\delta/(4t) + \cos\phi_x+1}{(t_z/t)\sin\phi_z}\right|\leq 1}$ (condition I);

\item $\displaystyle{\left(0,\frac{\pi}{a},k_w\right)}$ and $\displaystyle{\left(0,\frac{\pi}{a},\frac{\pi}{a}-k_w\right)}$, where $k_wa=\sin^{-1}c_{2}$ if $\displaystyle{|c_{2}|\equiv\left|\frac{\delta/(4t) + \cos\phi_x-1}{(t_z/t)\sin\phi_z}\right|\leq 1}$ (condition II);

\item $\displaystyle{\left(\frac{\pi}{a},0,k_w\right)}$ and $\displaystyle{\left(\frac{\pi}{a},0,\frac{\pi}{a}-k_w\right)}$, where $k_wa=\sin^{-1}c_{3}$ if $\displaystyle{|c_{3}|\equiv\left|\frac{\delta/(4t) - \cos\phi_x+1}{(t_z/t)\sin\phi_z}\right|\leq 1}$ (condition III);

\item $\displaystyle{\left(\frac{\pi}{a},\frac{\pi}{a},k_w\right)}$ and $\displaystyle{\left(\frac{\pi}{a},\frac{\pi}{a},\frac{\pi}{a}-k_w\right)}$, where $k_wa=\sin^{-1}c_{4}$ if $\displaystyle{|c_{4}|\equiv\left|\frac{\delta/(4t) - \cos\phi_x-1}{(t_z/t)\sin\phi_z}\right|\leq 1}$ (condition IV);
\end{enumerate}
It can be seen that the gapless points appear in pairs if the corresponding condition is satisfied by tuning the control parameters $\delta/t$, $t_{z}/t$, and $\phi_{x}$. If fact, we may have one pair of gapless points if $|t_z\sin\phi_z/(t\cos\phi_x)|\leq1$, two pairs if $|t_z\sin\phi_z/t|\leq1$ and $|t_z\sin\phi_z/(t\cos\phi_x|)>1$, and four pairs if $|t_z\sin\phi_z/t|>1$. As demonstrated in the main text, these gapless points indeed correspond to the Weyl seminmetals for single particle spectra, which should be readily observed in contrast to the proposal for Rashba SO-coupled Fermi superfluids~\cite{Gong11,Xu14,YXusm15,Husm14}.

\subsection*{Effective 2D Hamiltonian in lattices} \label{APPB}
For convenience, we write down explicitly the Hamiltonian of the system in 2D lattices,
\begin{eqnarray}
H_{0}({\mathbf{k}})=\sum_{{\mathbf{k}},\sigma\sigma'} \hat{c}_{{\mathbf{k}}\sigma}^{\dag}\Big[\epsilon({\mathbf{k}})I +\sum_{\alpha=x,y,z}d_{\alpha}({\mathbf{k}})\hat{\sigma}_{\alpha}\Big]_{\sigma\sigma'}\hat{c}_{{\mathbf{k}}\sigma'},
\end{eqnarray}
where ${\mathbf{k}}=(k_{x},k_{y})$ is in the first Brillouin zone and
\begin{align}
\epsilon({\mathbf k}) &= 2 t \sin\phi_x\sin(k_x a),\nonumber
\\
d_x({\mathbf k}) &= 2 t_0\sin(k_y a), \nonumber
\\
d_y({\mathbf k}) &= -2 t_0 \sin(k_x a), \nonumber \\
d_z({\mathbf k}) &= -2 t [\cos\phi_x\cos(k_x a)+\cos(k_y a)] -\frac{\delta}{2}.
\end{align}

\section{Synthetic magnetic field}

\begin{figure}[ht]
\includegraphics[width=0.55\columnwidth]{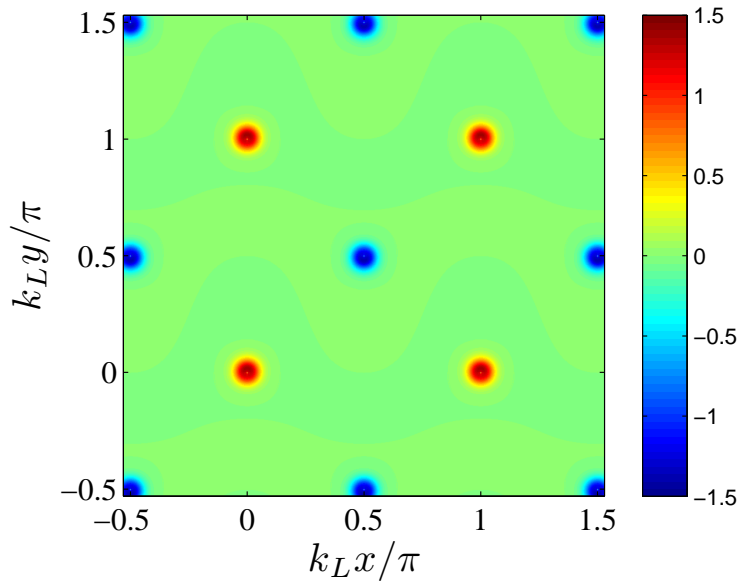}
\caption{(color online). Spatial distribution of the synthetic magnetic ${\mathbf B}_{\rm GF}$ (in units of Tesla) for $\delta=0.1 E_L$ and $\vartheta=\pi/2$. For optical lattices generated by the red-detuned (blue-detuned) lasers, atoms are trapped close to the the red (blue) sites.} \label{smf}
\end{figure}

The synthetic magnetic field can be easily obtained by following the standard procedure~\cite{Dalibard11,Goldman2014,Zhai15}. Specifically, we diagonalize the internal-state Hamiltonian ${\boldsymbol h}_{\rm in}$ for an arbitrary position ${\bf r}$. For simplicity, we consider only the flux lattice emerging on the $xy$ plane with fixed $\vartheta=\pi/2$. The resulting eigenstate that corresponds to the lower energy takes the form $|\chi_-({\bf r})\rangle = \begin{pmatrix}\cos\theta e^{-i\varphi} e^{i{\kappa_x x}}\\ \sin\theta\end{pmatrix}$, where $\cot\theta =(M_{x}^{2}+M_{y}^{2})^{1/2}/[-(M_{x}^{2}+M_{y}^{2}+\delta^2/4)^{1/2}+\delta/2]$ and $\tan\varphi = {M_y}/{M_x}$~\cite{Deng14}. Then, under the adiabatic approximation, the vector potentials ${\mathbf A}'$ can be straightforwardly evaluated as follows:
\begin{subequations}
\begin{align}
A_x' &=  \frac{i\hbar}{e} \langle\chi_-({\bf r})|{\partial_x}|\chi_-({\bf r}) \rangle= -\frac{\hbar}{e} \left(\kappa_x \cos^2
\theta - \cos^2\theta\frac{\partial \varphi}{\partial x}\right),  \\
A_y' &=  \frac{i\hbar}{e}\langle\chi_-({\bf r})|{\partial_y}|\chi_-({\bf r})\rangle = \frac{\hbar}{e} \cos^2\theta\frac{\partial \varphi}{\partial y},
\end{align}
\end{subequations}
where $e$ is the charge of the electron. Now the synthetic magnetic field is
\begin{align}
{\mathbf B}_{\rm GF}' &=  \nabla\times {\mathbf A}'= \left(\frac{\partial A_y'}{\partial x} -
\frac{\partial A_x'}{\partial y}\right)\hat{z}= -\frac{\hbar}{e} \sin (2\theta)\left[\kappa_x \frac{\partial \theta}{\partial y} + \left(\frac{\partial \theta}{\partial x}\frac{\partial
\varphi}{\partial y} - \frac{\partial \theta}{\partial
y}\frac{\partial \varphi}{\partial x}\right)\right]\hat{z}.
\end{align}

Figure~\ref{smf} shows the typical spatial distribution of the synthetic magnetic field which takes the form of  the optical flux lattices. At first sight, it may appear that the net optical flux vanishes. However, because the atoms are trapped only near the red (blue) sites for the optical lattices generated by the red-detuned (blue-detuned) lasers, the net magnetic flux for atomic gases is nonzero.

\section{Mean-field solution of the interacting model Hamiltonian}
\label{APPC}
The mean-field BdG Hamiltonian for our system with $s$-wave contact interaction takes the form
\begin{eqnarray}
{\mathcal{H}}_{\text{BdG}}(\mathbf{k})=\left(\begin{array}{cc}
{\mathcal{D}}({\bf k)} & \Delta_{{\bf {Q}}}I_{2\times2}\\
\Delta_{{\bf {Q}}}^{*}I_{2\times2} & -\hat{\sigma}_{y}{\mathcal{D}}^{*}({\bf -k})\hat{\sigma}_{y}
\end{array}\right),
\end{eqnarray}
where $I_{2\times2}$ is a $2\times2$ identity matrix and
\begin{eqnarray}
{\mathcal{D}}({\bf k})=\left(\begin{array}{cc}
\xi_{{\bf \frac{Q}{2}+k}}+d_{z}({\bf \frac{Q}{2}+k}) & d_{x}({\bf \frac{Q}{2}+k})-id_{y}({\bf \frac{Q}{2}+k})\\
d_{x}({\bf \frac{Q}{2}+k})+id_{y}({\bf \frac{Q}{2}+k}) & \xi_{{\bf \frac{Q}{2}+k}}-d_{z}({\bf \frac{Q}{2}+k})
\end{array}\right).
\end{eqnarray}
It can be shown that the Hamiltonian ${\mathcal{H}}_{\text{BdG}}({\bf k})$ has an inherent PH symmetry under the PH operator $\Lambda=\hat{\tau}_{y}\otimes\hat{\sigma}_{y}K$, where $\hat{\tau}_{y}$ is the Pauli matrix acting on the particle-hole space and $K$ is the complex conjugate operator.

In general, with the zero-temperature thermodynamic potential
\begin{align}
\Omega=-\frac{1}{U_{0}}|\Delta_{\mathbf Q}|^{2}+\frac{1}{\mathcal{S}}\sum_{{\bf k}}\xi_{{\bf k}}+\frac{1}{2\mathcal{S}}\sum_{\eta,{\bf k}}|E_{\eta,{\bf k}}^{\nu=+}|,
\end{align}
the system can be solved by seeking the lowest free energy $F=\Omega+\mu n$, which leads to the self-consistent saddle-point equations for the pairing gap $\Delta_{\mathbf Q}$, the particle density number $n$, and the FF momentum $\mathbf Q$, i.e.,
\begin{subequations}
\begin{align}
\frac{{\partial\Omega}}{{\partial\Delta_{\mathbf Q}}}&=0,\\
\frac{{\partial\Omega}}{{\partial\mu}}&=-n,\\
\frac{{\partial\Omega}}{{\partial {\mathbf Q}}}&=0.
\end{align}
\end{subequations}
This procedure, however, requires high-precision numerical differentiation, which is often difficult to achieve in practice.

Here, to avoid the numerical difficulty, we perform the derivatives in the saddle-point equations analytically via the Green's-function method. To this end, we introduce the field operators
\begin{subequations}
\begin{align}
\hat c_{{\bf \frac{Q}{2}+k,\sigma}}(\tau)&= \exp({H}\tau)\hat c_{{\bf \frac{Q}{2}+k,\sigma}}\exp(-{H}\tau),\\
\hat c_{{\bf \frac{Q}{2}-k,\sigma}}^{\dag}(\tau)&= \exp({H}\tau)\hat c_{{\bf \frac{Q}{2}-k,\sigma}}^{\dag}\exp(-{H}\tau),
\end{align}
\end{subequations}
with $\tau=it$ being the imaginary time. In terms of the Nambu representation, the Heisenberg equations for these operators take the form
\begin{align}
 & \frac{\partial\hat{\Psi}_{{\bf k,Q}}}{\partial\tau}=\left[{H},\hat{\Psi}_{{\bf k,Q}}(\tau)\right]=-\mathcal{H}_{\text{BdG}}({\bf k})\hat{\Psi}_{{\bf k,Q}}(\tau),
\end{align}
where, as in the main text, $\hat{\Psi}_{{\bf {k,Q}}}=\left(\hat{c}_{{\bf {\frac{Q}{2}+k},\uparrow}},\hat{c}_{{\bf{\frac{Q}{2}+k},\downarrow}},\hat{c}_{{\bf {\frac{Q}{2}-k},\downarrow}}^{\dagger},-\hat{c}_{{\bf {\frac{Q}{2}-k},\uparrow}}^{\dagger}\right)^{T}$ is the Nambu operator. Next, we define the single-particle Green's function as
\begin{align}
\mathcal{G}({\bf k},\tau)=-\langle\mathcal{T}_{\tau}\hat{\Psi}_{{\bf k,Q}}(\tau)\hat{\Psi}_{{\bf k,Q}}^{\dag}\rangle,
\end{align}
where $\mathcal{T}_{\tau}$ is the time-ordering operator. The Fourier transform of the Green's function is then
\begin{align}
\mathcal{G}({\bf k},\omega_{n}) & =\frac{I_{4\times4}}{i\omega_{n}-\mathcal{H}_{\text{BdG}}({\bf k})},
\end{align}
where $\omega_{n}=(2n+1)\pi/\beta$ is the Matsubara frequency for fermions; $\beta=1/(k_{B}T)$ is the inverse temperature, with $k_{B}$ being the Boltzmann constant and $T$ being the temperature; and $I_{4\times4}$ is a $4\times 4$ identity matrix. In particular, the usual and ``anomalous'' Green's functions can be constructed as
\begin{subequations}
\begin{align}
G_{\uparrow}({\bf k},\tau) & =\mathcal{G}_{11}({\bf k},\tau)=-\langle\mathcal{T}_{\tau}\hat c_{{\bf \frac{Q}{2}+k},\uparrow}(\tau)\hat c_{{\bf \frac{Q}{2}+k},\uparrow}^{\dag}\rangle,\\
G_{\downarrow}({\bf k},\tau) & =\mathcal{G}_{22}({\bf k},\tau)=-\langle\mathcal{T}_{\tau}\hat c_{{\bf \frac{Q}{2}+k},\downarrow}(\tau)\hat c_{{\bf \frac{Q}{2}+k},\downarrow}^{\dag}\rangle,\\
F({\bf k},\tau) &=\mathcal{G}_{13}({\bf k},\tau) =-\langle\mathcal{T}_{\tau}\hat c_{{\bf \frac{Q}{2}+k},\uparrow}(\tau)\hat c_{{\bf \frac{Q}{2}-k},\downarrow}\rangle.
\end{align}
\end{subequations}
The self-consistent equations of $\Delta_{{\bf Q}}$ and $n$ can then be evaluated to yield
\begin{align}
\Delta_{\bf Q} &= \frac{U_0}{\mathcal{S}}\sum_{\mathbf{k}}F({\bf
k},0^-)=\frac{U_0}{\mathcal{S}}\sum_{{\bf k}}\left[f(\mathcal{H}_{\text{BdG}}({\bf k}))\right]_{13},\label{edelq}\\
n &= \frac{1}{\mathcal{S}}\sum_{\mathbf{k}}\left[G_{\uparrow}({\bf k},0^-)+G_{\downarrow}({\bf k},0^-) \right] =\frac{1}{\mathcal{S}}\sum_{\bf k}\sum_{\sigma=1,2}
\left[f(\mathcal{H}_{\text{BdG}}({\bf k}))\right]_{\sigma\sigma},\label{en}
\end{align}
where $f(x) = \frac{1}{e^{\beta x}+1} =
\frac{1}{\beta}\sum_{n=-\infty}^{\infty} \frac{e^{i\omega_n
0^+}}{i\omega_n-x}$ is the Fermi-Dirac distribution. Moreover, to obtain the saddle-point equation for ${\bf Q}$, we make use of the functional path-integral formalism for the thermodynamic potential
\begin{align}
e^{-\beta{\mathcal S}\Omega} & =\int D[c_{{\bf k},\sigma},c_{{\bf k},\sigma}^{\dagger}]\exp\left[-\frac{1}{2}\int_{0}^{\beta}d\tau\sum_{{\bf k}}\Psi_{{\bf k}}^{\dagger}[\partial_{\tau}+\mathcal{H}_{{\rm {BdG}}}({\bf k})]\Psi_{{\bf k}}+\beta\mathcal{S}\frac{\Delta_{{\bf Q}}^{2}}{U_{0}}\right],
\end{align}
which leads to
\begin{eqnarray}
\frac{\partial \Omega}{\partial {\bf Q}}=\frac{1}{2{\mathcal S}}\sum_{{\bf k}}{\rm Tr}\left[\frac{\partial\mathcal{H}_{{\rm {BdG}}}({\bf k})}{\partial{\bf Q}}f(\mathcal{H}_{{\rm {BdG}}}({\bf k}))\right]=0.\label{eoq}
\end{eqnarray}
Equations (\ref{edelq}), (\ref{en}), and (\ref{eoq}) then form a closed set of equations which allows us to solve for $\Delta_{\mathbf Q}$, $n$, and ${\mathbf Q}$. Since $\partial \mathcal{H}_{\mathrm{BdG}}(\mathbf{k})/{\partial {\mathbf Q}}$ and $f(\mathcal{H}_{\mathrm{BdG}}(\mathbf{k}))$ are $4\times 4$ matrices, the numerical method for solving this set of nonlinear equations is very efficient at the zero temperature.

Finally, let us comment on the validity of the mean-field approach to the superfluid phase diagram. It is known from previous studies that mean-field theory predicts various qualitative features of 1D and 2D interacting quantum gases in the weakly interacting regime. For example, even in the most serious 1D cases, the qualitative mean-field predictions of topological effects, including the existence of Majorana fermions~\cite{Lutchyn10,Oreg10} and dark solitons~\cite{Pustilnik15}, are not invalidated by strong quantum fluctuations in 1D~\cite{Fidkowski11,Sau11,Delande14}.

\end{widetext}


\begin{thebibliography}{10}
\bibitem{Xiao10} D. Xiao, M.-C. Chang, and Q. Niu, Rev. Mod. Phys. \textbf{82}, 1959 (2010).

\bibitem{Hasan10} M. Z. Hasan and C. L. Kane, Rev. Mod. Phys. \textbf{82}, 3045 (2010).

\bibitem{Qi11} X.-L. Qi and S.-C. Zhang, Rev. Mod. Phys. \textbf{83}, 1057 (2011).

\bibitem{Kane05} C.L. Kane and E.J. Mele, Phys. Rev. Lett. {\bf 95}, 146802 (2005).

\bibitem{Bernevig06} B.A. Bernevig, T.L. Hughes, and S.-C. Zhang, Science {\bf 314}, 1757 (2006).

\bibitem{Hsieh08} D. Hsieh, D. Qian, L. Wray, Y. Xia, Y. S. Hor, R.J. Cava, and M. Z. Hasan, Nature (London) {\bf 452}, 970 (2008).

\bibitem{Mourik12} V. Mourik, K. Zuo, S.M. Frolov, S.R. Plissard, E.P.A.M. Bakkers, and L.P. Kouwenhoven, Science {\bf 336}, 1003 (2012).

\bibitem{MTDeng} M.T. Deng, C.L. Yu, G.Y. Huang, M. Larsson, P. Caroff, and H.Q. Xu, Nano Lett {\bf 12}, 6414 (2012).

\bibitem{Das12} A. Das, Y. Ronen, Y. Most, Y. Oreg, M. Heiblum, and H. Shtrikman, Nat. Phys. {\bf 8}, 887 (2012).

\bibitem{Xusm15} S.-Y. Xu, I. Belopolski, N. Alidoust, M. Neupane, G. Bian, C. Zhang, R. Sankar, G. Chang, Z. Yuan, C.-C. Lee, S.-M. Huang, H. Zheng, J. Ma, D.S. Sanchez, B. Wang, A. Bansil, F. Chou, P.P. Shibayev, H. Lin, S. Jia, M.Z. Hasan, Science \textbf{349}, 613 (2015).

\bibitem{Lusm15} L. Lu, Z. Wang, D. Ye, L. Ran, L. Fu, J.D. Joannopoulos, M. Solja\u{c}i\'{c}, Science \textbf{349}, 622 (2015).

\bibitem{Lvsm15} B.Q. Lv, H.M. Weng, B.B. Fu, X.P. Wang, H. Miao, J. Ma, P. Richard, X.C. Huang, L.X. Zhao, G.F. Chen, Z. Fang, X. Dai, T. Qian, and H. Ding, Phys. Rev. X \textbf{5}, 031013 (2015).

\bibitem{Soluyanovsm15} A.A. Soluyanov, D. Gresch, Z. Wang, Q.S. Wu, M. Troyer, X. Dai, B.A. Bernevig, Nature (London) \textbf{527}, 495 (2015).

\bibitem{Lin11} Y.-J. Lin, K. Jim\'{e}nez-Garc\'{i}a, and I.B. Spielman, Nature (London) {\bf 471}, 83 (2011).

\bibitem{Wang12} P. Wang, Z.-Q. Yu, Z. Fu, J. Miao, L. Huang, S. Chai, H. Zhai, and J. Zhang, Phys. Rev. Lett. {\bf 109}, 095301 (2012).

\bibitem{Cheuk12} L.W. Cheuk, A.T. Sommer, Z. Hadzibabic, T. Yefsah, W.S. Bakr, and M.W. Zwierlein, Phys. Rev. Lett. {\bf 109}, 095302 (2012).

\bibitem{zjy12} J.-Y. Zhang, S.-C. Ji, Z. Chen, L. Zhang, Z.-D. Du, B. Yan, G.-S. Pan, B. Zhao, Y.-J. Deng, H. Zhai, S. Chen, and J.-W. Pan, Phys. Rev. Lett. {\bf 109}, 115301 (2012).

\bibitem{Dalibard11} J. Dalibard, F. Gerbier, G. Juzeli\={u}nas, and P. \"{O}hberg, Rev. Mod. Phys.\textbf{ 83}, 1523 (2011).

\bibitem{Goldman2014} N. Goldman, G. Juzeli\={u}nas, P. \"{O} hberg, and I.B Spielman, Rep. Prog. Phys. {\bf77}, 126401 (2014).

\bibitem{Zhai15} H. Zhai, Rep. Prog. Phys. {\bf78}, 026001 (2015).

\bibitem{Sinha11} S. Sinha, R. Nath, and L. Santos, Phys. Rev. Lett. {\bf 107}, 270401 (2011).

\bibitem{Deng12} Y. Deng, J. Cheng, H. Jing, C.-P. Sun, and S. Yi, Phys. Rev. Lett. {\bf 108}, 125301 (2012).

\bibitem{Sato09} M. Sato, Y. Takahashi, and S. Fujimoto, Phys. Rev. Lett. {\bf 103}, 020401 (2009).

\bibitem{Jiang11} L. Jiang, T. Kitagawa, J. Alicea, A.R. Akhmerov, D. Pekker, G. Refael, J.I. Cirac, E. Demler, M.D. Lukin, and P. Zoller, Phys. Rev. Lett. {\bf 106}, 220402 (2011).

\bibitem{Gong11} M. Gong, S. Tewari, and C. Zhang, Phys. Rev. Lett. {\bf 107}, 195303 (2011).

\bibitem{Hu11} H. Hu, L. Jiang, X.-J. Liu, and H. Pu, Phys. Rev. Lett. {\bf 107}, 195304 (2011).

\bibitem{Zhai11} Z.-Q. Yu and H. Zhai, Phys. Rev. Lett. {\bf 107}, 195305 (2011).

\bibitem{Xu11} Z.F. Xu, R. L\"{u}, and L. You, Phys. Rev. A {\bf 83}, 053602 (2011).

\bibitem{Atala13} M. Atala, M. Aidelsburger, J.T. Barreiro, D. Abanin, T. Kitagawa, E. Demler, and I. Bloch, Nat. Phys. {\bf 9}, 795 (2013).

\bibitem{Aidelsburger13} M. Aidelsburger, M. Atala, M. Lohse, J.T. Barreiro, B. Paredes, and I. Bloch, Phys. Rev. Lett. {\bf111}, 185301 (2013).

\bibitem{Miyake13} H. Miyake, G.A. Siviloglou, C.J. Kennedy, W.C. Burton, and W. Ketterle, Phys. Rev. Lett. {\bf111}, 185302 (2013).

\bibitem{Jotzu14} G. Jotzu, M. Messer, R. Desbuquois, M. Lebrat, T. Uehlinger, D. Greif, and T. Esslinger, Nature (London) {\bf515}, 237 (2014).

\bibitem{Mancini15} M. Mancini, G. Pagano, G. Cappellini, L. Livi, M. Rider, J. Catani, C. Sias, P. Zoller, M. Inguscio, M. Dalmonte, and L. Fallani, Science {\bf349}, 1510 (2015).

 \bibitem{Stuhl15} B.K. Stuhl, H.-I. Lu, L.M. Aycock, D. Genkina, and I.B. Spielman, Science {\bf349}, 1514 (2015).

\bibitem{Nakajima16} S. Nakajima, T. Tomita, S. Taie, T. Ichinose1, H. Ozawa, L. Wang, M. Troyer, and Y. Takahashi1, Nat. Phys. {\bf 12}, 296 (2016).

\bibitem{Lohse16} M. Lohse, C. Schweizer, O. Zilberberg, M. Aidelsburger, and I. Bloch, Nat. Phys. {\bf 12}, 350 (2016).

\bibitem{Qu13} C. Qu, Z. Zheng, M. Gong, Y. Xu, L. Mao, X. Zou, G. Guo, and C. Zhang, Nat. Commun. {\bf4}, 2710 (2013).

\bibitem{Zhang13} W. Zhang and W. Yi, Nat. Commun. {\bf4}, 2711 (2013).

\bibitem{Xu14} Y. Xu, R.-L. Chu, and C. Zhang, Phys. Rev. Lett. {\bf112}, 136402 (2014).

\bibitem{Cao14} Y. Cao, S.-H. Zou, X.-J. Liu, S. Yi, G.-L. Long, and H. Hu, Phys. Rev. Lett. {\bf113}, 115302 (2014).

\bibitem{Ruseckas05} J. Ruseckas, G. Juzeli\={u}nas, P. \"{O}hberg, and M. Fleischhauer, Phys. Rev. Lett. {\bf95}, 010404 (2005).

\bibitem{Vaishnav08} J.Y. Vaishnav and C.W. Clark, Phys. Rev. Lett. {\bf100}, 153002 (2008).

\bibitem{Dalibard10} G. Juzeli\={u}nas, J. Ruseckas, and J. Dalibard, Phys. Rev. A {\bf81}, 053403 (2010).

\bibitem{Campbell11} D.L. Campbell, G. Juzeli\={u}nas, and I. B. Spielman, Phys. Rev. A {\bf84}, 025602 (2011).

\bibitem{Anderson12} B.M. Anderson, G. Juzeli\={u}nas, V.M. Galitski, and I.B. Spielman, Phys. Rev. Lett. {\bf108}, 235301 (2012).

\bibitem{Huang15} L. Huang, Z. Meng, P. Wang, P. Peng, S.-L. Zhang, L. Chen, D. Li, Q. Zhou, and J. Zhang, Nat. Phys. {\bf12}, 540 (2016).

\bibitem{Wu15} Z. Wu, L. Zhang, W. Sun, X.-T. Xu, B.-Z. Wang, S.-C. Ji, Y. Deng, S. Chen, X.-J. Liu, and J.-W. Pan, Science {\bf354}, 83 (2016).


\bibitem{Cooper11}  N.R. Cooper, Phys. Rev. Lett. {\bf106}, 175301 (2011).

\bibitem{Deng14}  Y. Deng, J. Cheng, H. Jing, and S. Yi, Phys. Rev. Lett. {\bf112}, 143007 (2014).

\bibitem{Liu14} X.-J. Liu, K.T. Law, and T.K. Ng, Phys. Rev. Lett. {\bf112}, 086401 (2014); {\bf113}, 059901 (2014).

\bibitem{Xusm16} Y. Xu and L.-M. Duan, Phys. Rev. A {\bf94}, 053619 (2016).

\bibitem{Crosse14} J. A. Crosse, Phys. Rev. B {\bf90}, 235403 (2014).

\bibitem{AZ} A. Altland and M. R. Zirnbauer, Phys. Rev. B {\bf55}, 1142 (1997).

\bibitem{Ludwig08} A. P. Schnyder, S. Ryu, A. Furusaki, and A. W. W. Ludwig, Phys. Rev. B {\bf78}, 195125 (2008).


\bibitem{TKNN} D.J. Thouless, M. Kohmoto, M.P. Nightingale, and M. den Nijs, Phys. Rev. Lett. {\bf49}, 405 (1982).

\bibitem{gFF-I} We have checked that there are no edge modes if we impose the hard-wall confinement along the $x$ direction. The bulk-edge correspondence theorem is therefore not violated.

\bibitem{Aidelsburger15} M. Aidelsburger, M. Lohse, C. Schweizer, M. Atala, J.T. Barreiro, S. Nascimb\`{e}ne, N.R. Cooper, I. Bloch, and N. Goldman, Nat. Phys. {\bf11}, 162 (2015).

\bibitem{Duca15} L. Duca, T. Li, M. Reitter, I. Bloch, M. Schleier-Smith, and U. Schneider, Science {\bf347}, 288 (2015).

\bibitem{Li16} T. Li, L. Duca, M. Reitter, F. Grusdt, E. Demler, M. Endres, M. Schleier-Smith, I. Bloch, and U. Schneider, Science {\bf352}, 1094 (2016).

\bibitem{Flaschner16} N. Fl\"{a}schner, B.S. Rem, M. Tarnowski, D. Vogel, D.-S. L\"{u}hmann, K. Sengstock, and C. Weitenberg, Science {\bf352}, 1091 (2016).

\bibitem{YXusm15} Y. Xu, F. Zhang, and C. Zhang, Phys. Rev. Lett. \textbf{115}, 265304 (2015).

 \bibitem{Husm14} H. Hu, L. Dong, Y. Cao, H. Pu, and X.-J. Liu, Phys. Rev. A \textbf{90}, 033624 (2014).

\bibitem{Lutchyn10} R.M. Lutchyn, J.D. Sau, and S.D. Sarma, Phys. Rev. Lett. \textbf{105}, 077001 (2010).

\bibitem{Oreg10} Y. Oreg, G. Refael, and F. von Oppen, Phys. Rev. Lett. \textbf{105}, 177002 (2010).

\bibitem{Pustilnik15} M. Pustilnik and K.A. Matveev, Phys. Rev. B \textbf{92}, 195146 (2015).

\bibitem{Fidkowski11} L. Fidkowski, R.M. Lutchyn, C. Nayak, and M.P.A. Fisher, Phys. Rev. B \textbf{84}, 195436 (2011).

\bibitem{Sau11} J.D. Sau, B.I. Halperin, K. Flensberg, and S.D. Sarma, Phys. Rev. B \textbf{84}, 144509 (2011).

\bibitem{Delande14} D. Delande and K. Sacha, Phys. Rev. Lett. \textbf{112}, 040402 (2014).

\end{thebibliography}
\end{document}